\documentclass[journal]{IEEEtran}[11pt]
\ifCLASSINFOpdf
\usepackage[pdftex]{graphicx}
\graphicspath{{../pdf/}{../jpeg/}}
\DeclareGraphicsExtensions{.pdf,.jpeg,.png}
\else
\usepackage[dvips]{graphicx}
\graphicspath{{../eps/}}
\DeclareGraphicsExtensions{.eps}
\fi\usepackage{graphicx}

\usepackage{subfloat}
\usepackage{graphics}
\usepackage{epsfig}
\usepackage{epstopdf}
\usepackage{stfloats}
\usepackage[cmex10]{amsmath}
\usepackage{algorithmic}
\usepackage{array}
\usepackage{mdwmath}
\usepackage{mdwtab}
\usepackage{graphicx}
\usepackage{subfigure}
\usepackage{color}
\usepackage{amsfonts,amssymb}
\usepackage{hyperref}
\hypersetup{hypertex = true,
	colorlinks = true, 
	linkcolor = blue,
	anchorcolor = blue,
	citecolor = blue 
}
\usepackage{multirow}
\usepackage{diagbox}
\usepackage{caption}
\usepackage{makecell}
\usepackage{lineno} 
\usepackage{mathrsfs}
\usepackage{booktabs}
\usepackage{makecell}
\usepackage{tabu} 
\usepackage{multirow}
\usepackage{multicol} 
\usepackage{multirow} 
\usepackage{float} 
\usepackage{makecell} 
\usepackage{booktabs}
\usepackage{amssymb} 
\usepackage{bbding}
\usepackage[linesnumbered,ruled]{algorithm2e}
\usepackage{amsfonts,amsthm,array} 
\usepackage[ruled]{algorithm2e}

\usepackage{threeparttable} 

\begin{document}
	
	
	\title{Secure Offloading in NOMA-Aided Aerial MEC Systems Based on Deep Reinforcement Learning \thanks{Manuscript received.}}
	
	\author{Hongjiang~Lei, 
		Mingxu~Yang,
		Jiacheng~Jiang,
		Ki-Hong~Park, 
		and
		Gaofeng~Pan 
		\thanks{This work was supported by the National Natural Science Foundation of China under Grant 62171031 and 61971080. (Corresponding author: \textit{Jiacheng~Jiang, Gaofeng~Pan}.)}
		\thanks{Hongjiang~Lei, Mingxu~Yang, and Jiacheng~Jiang are with the School of Communications and Information Engineering, Chongqing University of Posts and Telecommunications, Chongqing 400065, China (e-mail: leihj@cqupt.edu.cn, ymx13648245248@163.com, cquptjjc@163.com).}
		\thanks{Ki-Hong~Park is with the CEMSE Division, King Abdullah University of Science and Technology (KAUST), Thuwal 23955-6900, Saudi Arabia (e-mail: kihong.park@kaust.edu.sa).}
		\thanks{Gaofeng~Pan is with the School of Cyberspace Science and Technology, Beijing Institute of Technology, Beijing 100081, China (e-mail: gaofeng.pan.cn@ieee.org).}
	}
\maketitle
\begin{abstract}
	
Mobile edge computing (MEC) technology can reduce user latency and energy consumption by offloading computationally intensive tasks to the edge servers. Unmanned aerial vehicles (UAVs) and non-orthogonal multiple access (NOMA) technology enable the MEC networks to provide offloaded computing services for massively accessed terrestrial users conveniently. However, the broadcast nature of signal propagation in NOMA-based UAV-MEC networks makes it vulnerable to eavesdropping by malicious eavesdroppers. In this work, a secure offload scheme is proposed for NOMA-based UAV-MEC systems with the existence of an aerial eavesdropper. The long-term average network computational cost is minimized by jointly designing the UAV's trajectory, the terrestrial users' transmit power, and computational frequency while ensuring the security of users' offloaded data. Due to the eavesdropper's location uncertainty, the worst-case security scenario is considered through the estimated eavesdropping range. Due to the high-dimensional continuous action space, the deep deterministic policy gradient algorithm is utilized to solve the non-convex optimization problem. Simulation results validate the effectiveness of the proposed scheme.

\end{abstract}


\section{Introduction}
\label{sec:int}

\subsection{Background and Related Works}

\IEEEPARstart{M}{obile} edge computing (MEC) enables offloading computation and storage resources to the network edge, allowing for task processing closer to where data is generated, thereby reducing data transmission latency and bandwidth requirements on Internet of Things (IoT) systems \cite{AbbasN2018IOTJ}-\cite{YeY2022TWC}.
However, there remains a challenging issue for terminal devices located in remote areas or mountainous regions, as they need more infrastructure coverage to access reliable MEC servers.
In such circumstances, due to the absence of MEC support and infrastructure limitations, terminal devices face significant restrictions and limitations in terms of computational and communication capabilities.
Through the establishment of line of sight (LoS) connections with devices used by ground users (GUs), unmanned aerial vehicles (UAVs) can serve as ``aerial MEC servers", offering significant offloading services characterized by minimal network overhead and execution latency \cite{ZengY2016CM}.
In Ref. \cite{HuQ2019IOT}, the total delay of all users was reduced by optimizing the scheduling variables, task offload ratio, and trajectory of the UAV. Furthermore, a novel penalty dual decomposition (PDD) algorithm was developed using the augmented Lagrangian method to solve the non-convex problem with mixed integers.
The energy consumption of the UAV was reduced by optimizing the CPU frequencies, the amount of offloading, the transmission power, and the UAV's trajectory. To address the non-convex problem, Liu \textit{et al}. proposed the successive convex approximation (SCA)-based and the decomposition and iteration-based algorithms in their work \cite{LiuY2020IOT}.
The energy efficiency was also minimized in the scenarios where the UAV operated in both 1D and 2D trajectories \cite{SunC2021TWC}.
In Ref. \cite{LiuB2022TVT}, the MEC system's energy consumption was minimized by simultaneously optimizing the UAV’s beamforming vectors, CPU frequency, trajectory, and the GUs’ transmission power. An iterative algorithm featuring three stages was introduced to address the intricacies of the resultant non-convex optimization problem, each stage is solved by SCA, the semi-definite programming (SDP) and alternating optimization (AO) algorithms respectively.
In Ref. \cite{XuY2021TWC}, the authors maximized the weighted computation efficiency by concurrently optimizing task assignment, UAV's trajectory, CPU frequency, and power allocation and proposed an alternative algorithm based on the Dinkelbach method, Lagrange duality, and SCA technique.
{Ref. \cite{HuZ2023TVT} investigated the 3D trajectory and resource allocation optimization scheme in space-air-ground integrated networks, in which UAVs acted as MEC servers to provide computing services to GUs, and tasks that the UAVs cannot process were offloaded to the satellite for computation.} 

Nonorthogonal multiple access (NOMA) offers the advantage of improved spectral efficiency and increased system capacity by allowing multiple users to share the same time-frequency resources and has been extensively used in aerial MEC networks to reduce the latency and energy cost \cite{JiJ2021IOT}-\cite{FengW2021SAC}. 
In Ref. \cite{JiJ2021IOT}, the weighted energy consumptions of the aerial MEC systems with orthogonal multiple access (OMA) and NOMA schemes were minimized through the joint optimization of the computation resource allocation and UAV trajectory. 
In Ref. \cite{BudhirajaI2021SJ}, the total energy consumption was minimized by joint designing the task allocation, time  scheduling, transmit power, and UAV trajectory. The nonlinear optimization problem was divided into two subproblems and was solved by the iterative algorithm.
Zhang \textit{et al}. \cite{ZhangX2020IOT} studied the energy-efficient multiple UAVs-assisted MEC networks and minimized the weighted system energy consumption by jointly optimizing the radio and computation resources.
In Ref. \cite{FengW2021SAC}, the hybrid beamforming and NOMA techniques were utilized in the UAV-MEC networks
to enhance the computation performance.
The total computation rate of all the GUs was maximized by optimizing the UAV’s three-dimensional (3D) position, hybrid beamforming design, and computational resource allocation. Two novel algorithms were introduced to deal with the non-convex optimization challenges for partial and binary offloading scenarios.

The broadcast nature of wireless channels allows eavesdroppers to also benefit from the LoS link provided by the UAV-MEC communication system, which makes user offloading information easy to intercept by terrestrial eavesdroppers \cite{XuY2021TCOM}-\cite{MaoW2023TWC} and aerial eavesdroppers \cite{ZhouY2020TCOM}-\cite{LuW2022TCOM}.
Authors in Ref. \cite{XuY2021TCOM} studied the secrecy performance of a dual-UAV-assisted MEC system with time division multiple access (TDMA) and NOMA schemes, wherein one UAV was utilized as the ``aerial MEC server" and the other utilized as the friendly jammer to fight with the terrestrial eavesdroppers.
The minimum secrecy computing capacity was maximized by jointly optimizing computation resources and UAVs' trajectories and
an algorithm based on block coordinate descent (BCD) was introduced to handle the formulated problem.
Ref. \cite{ChenP2023TIV} investigated the security performance of a UAV-MEC system in the presence of multiple location-uncertain terrestrial eavesdroppers, where the aerial user offloaded data to multiple terrestrial MEC servers.
The worst average secrecy capacity was maximized by jointly designing the UAV trajectory and transmit power, the power of artificial noise (AN) and the local computation ratio.
Mao \textit{et al.} investigated the robustness and security of multi-antenna aerial MEC networks in \cite{MaoW2023TWC}.
Considering the imperfect channel state information (CSI) during information offloading, the energy consumption of a multi-antenna UAV-assisted MEC network was minimized by jointly optimizing offloading time, beamforming vectors, CPU frequency, UAV's trajectory, and transmit power.
Zhou \textit{et al.} \cite{ZhouY2020TCOM} investigated the security, latency, and offloading performance of the aerial MEC systems and this study revealed a tradeoff between the security and execution delay for the aerial MEC systems.
To weaken the eavesdropping capability of flying eavesdroppers, the terrestrial jammer was employed to enhance the security  of aerial MEC systems with TDMA scheme \cite{LuW2021TII}, \cite{DingY2023TVT}, and NOMA scheme \cite{LuW2022TCOM}.
The minimum average secrecy capacity was maximized by optimizing the resources and trajectory of the aerial MEC server, respectively.

Considering the intricacy of task scheduling and resource allocation in UAV-MEC networks and the uncertainty of the environment, traditional optimization methods based on the SCA and BCD technologies are generally complicated to solve the formulated non-convex problems.
Deep reinforcement learning (DRL) has been widely applied to solve dynamic decision-making problems in recent years \cite{YangH2024TWC}-\cite{LinN2023TWC}.
In \cite{WangL2021TMC}, the consumed energy of all GUs was minimized by optimizing the UAVs’ trajectory, user scheduling, and computation resources.
An algorithm formulated on convex optimization principles and another algorithm based on DRL were proposed to address it iteratively and facilitate fast decision-making.
{To guarantee load fairness among UAVs, a 3D multi-UAV trajectory optimization scheme was investigated in Ref. \cite{HeY2023IOT}, wherein the GU selected the optimal UAV for task offloading and the formulated problem was solved by utilizing the deep deterministic policy gradient (DDPG) algorithm.}
Ref. \cite{ZhaoR2021TGCN} proposed a novel optimization framework wherein the latency, energy, and price in different combinations according to users' requirements, and a deep Q-network (DQN) based algorithm was proposed to deal with the offloading optimization problem. 
A secure model for transmission in the multiple UAV-MEC network was considered in \cite{LuW2022TNSE}, wherein the spiral placement algorithm was utilized to minimize the required server UAVs for full GUs coverage and two secure transmission schemes based on reinforcement learning were proposed to maximize the weighted delay and the residual energy utility. 
{TABLE \ref{table1} presents typical works related to aerial MEC systems}.

\begin{table*}
	\centering
	{
	\caption{Recent Works on Aerial MEC Systems}
	\label{table1}
	\begin{threeparttable}
	\begin{tabular}{c|c|p{1.3cm}<{\centering}|c|c|c|c|c}
		\Xhline{1.2pt}
		\textbf{Reference} & \textbf{Optimization Objective} & \textbf{\makecell[c]{UAV \\ Trajectory}} & \textbf{\makecell[c]{MA \\Scheme}} & \textbf{PLS} & \textbf{\makecell[c]{Aerial $E$}} & \textbf{\makecell[c]{Imperfect \\Location of $E$}} & \textbf{Method} \\
		\hline
		\cite{HuQ2019IOT}                  & Delay                               & 2D         & TDMA   &              &              &                & PDD  \\
		\hline
		\cite{LiuY2020IOT, SunC2021TWC}    & UAV consumed energy              & 2D         & TDMA   &              &              &                & SCA  \\
		\hline
		\cite{LiuB2022TVT}                 & User and UAV consumed energy     & 2D         & TDMA   &              &              &                & SCA, SDP, AO   \\
		\hline
		\cite{XuY2021TWC}                  & Energy efficiency                   & 2D         & FDMA   &              &              &                & Dinkelbach, SCA   \\
		\hline
		\cite{HuZ2023TVT}                  & Energy efficiency                   & 3D         & TDMA   &              &              &                & Dinkelbach, BCD  \\
		\hline
		\cite{ZhangX2020IOT}               & User and UAV consumed energy     & 2D         & NOMA   &              &              &                & SCA  \\
		\hline
		\cite{XuY2021TCOM}                 & Secrecy capacity                   & 2D         & NOMA   & $\checkmark$ &              &                & BCD     \\
		\hline
		\cite{MaoW2023TWC}                 & User and UAV consumed energy     & 2D         & TDMA   & $\checkmark$ &              &                & SCA      \\
		\hline
		\cite{ZhouY2020TCOM}               & Secrecy capacity                   & Fixed\tnote{a}      & NOMA   & $\checkmark$ & $\checkmark$ & $\checkmark$   & BCD, SCA      \\
		\hline
		\cite{LuW2021TII}                  & Secrecy capacity                   &  2D        & TDMA   & $\checkmark$ & $\checkmark$ &                & BCD, SCA     \\
		\hline
		\cite{DingY2023TVT}                & Secrecy capacity                   & 2D         & TDMA  & $\checkmark$ & $\checkmark$ & & Double DQN \\
		\hline  
		\cite{LuW2022TCOM}                 & Secrecy capacity                   &  2D        & NOMA   & $\checkmark$ & $\checkmark$ & $\checkmark$   & BCD, SCA   \\
		\hline
		\cite{LinN2023TWC}                 & Energy efficiency                   & 3D         & TDMA   &              &              &                & DQN  \\
		\hline
		\cite{WangL2021TMC}                & User consumed energy             &  3D        & FDMA   &              &              &                & BCD, DDPG  \\
		\hline
		\cite{HeY2023IOT}                  & User and UAV consumed energy      & 3D         & TDMA   &              &              &                & DDPG  \\
		\hline
		\cite{ZhaoR2021TGCN}               & User computing cost                & Fixed\tnote{a}      & FDMA   & $\checkmark$ & $\checkmark$ &                & DQN  \\
		\hline
		\cite{LuW2022TNSE}                 & System utility                  & Fixed\tnote{a}     & TDMA   & $\checkmark$ & $\checkmark$ &                & Nash Q-learning  \\
		\hline
		Our work                           & User delay and consumed energy   &  3D        & NOMA   & $\checkmark$ & $\checkmark$ & $\checkmark$   & DDPG  \\
		\Xhline{1.2pt}
	\end{tabular}
	\begin{tablenotes}
		\footnotesize
		\item Fixed\tnote{a}: The UAV hovers in a fixed position.
	\end{tablenotes}
\end{threeparttable}}
\end{table*}

\subsection{Motivation and Contributions}

Although the UAV-MEC system has been extensively investigated, the in-depth research and analysis in this field still need to be improved.
The scenarios involving location-uncertain eavesdropping and ensuring secure transmission of offloading data still need to be explored.
This work aims to ensure the security of user data transmission and computation efficiency while comprehensively considering eavesdropping location uncertainty.
The main contributions of this work can be summarised as follows.
\begin{enumerate}
	\item With the premise of ensuring secure offloading in a NOMA-aided aerial MEC system, we propose a new optimization scheme to minimize average computational cost by jointly designing the 3D trajectory of the UAV, the user's transmit power, and CPU frequency.
	To cope with the high-dimensional and continuous action space control problem in the UAV-MEC scenario, a DDPG-based trajectory and dynamic resource allocation (TDRA) scheme is proposed to solve the presented optimization challenge and obtain effective user resource allocation and UAV flight trajectory design.
	
	\item {Relative to \cite{ZhouY2020TCOM}, \cite{LuW2021TII}, \cite{LuW2022TCOM}, wherein 2D trajectory was jointly designed with transmit power and other parameters to obtain the security offloading in the UAV-MEC system with a location-uncertain eavesdropper. This work minimizes the average network computational cost by jointly designing the UAV's 3D trajectory, GUs' transmit power and computational frequency while ensuring the security of GUs' offloaded data. A DDPG-based TDRA scheme is proposed to solve the formulated problems.}
	
	\item {Relative to \cite{HuZ2023TVT}, \cite{LinN2023TWC}, \cite{WangL2021TMC}, \cite{HeY2023IOT} wherein 3D trajectory of the UAV was designed to optimize the energy efficiency of aerial MEC systems with OMA scheme, we design 3D trajectory of the UAV to minimize the average network computational cost of aerial MEC systems with NOMA scheme while considering security constraint.
		The worst-case security scenario with the eavesdropper's location uncertainty is considered through the estimated eavesdropping region. }
	
\end{enumerate}

\subsection{Organization}

The subsequent sections are structured as follows.
Section \ref{sec:sysmod} presents the system model and gives the corresponding optimization problem.
Section \ref{sec:DDPG} provides a detailed description of the DDPG-based TDRA scheme and gives the definitions related to DRL.
Section \ref{sec:Evaluation} provides simulation results and analysis.
Finally, the paper is summarized in Section \ref{sec:Conclusions}.
{ The notations and symbols used in this work are listed in TABLE \ref{table2}}.
\begin{table}[t]
	\centering
	\caption{Notation and Symbols}
	{
		\label{table2}
		\begin{tabular}{l|l}
			\Xhline{1.2pt}
			{Notation} & {Description} \\
			\hline
			${{\mathbf{q}}_S}\left( n \right)$         & Location of $S$ in the $n$th slot  \\
			\hline
			${{\mathbf{q}}_E}\left( n \right)$         & Location of $E$ in the $n$th slot  \\
			\hline
			${{\mathbf{\tilde q}}_E}$                  & Centre of $E$ estimated location  \\
			\hline
			${{\mathbf{w}}_J}$                         & Location of $J$ \\
			\hline
			${{\mathbf{w}}_{{U_k}}}\left( n \right)$   & Location of $U_k$ in the $n$th slot \\
			\hline
			${\delta _t}$                          & Time slot size\\
			\hline
			$v\left( n \right)$                    & The velocity of $S$\\
			\hline
			$\theta \left( n \right)$              & The polar angle of $S$\\
			\hline
			$\varphi \left( n \right)$             & The azimuthal angle of $S$\\
			\hline
			${p_k}\left( n \right)$, ${P_J}$       & Transmit power of $U_k$ and $J$\\
			\hline
			$P_k^{\max }$                          & Peak transmit power of $U_k$\\
			\hline
			$B$                                    & Channel bandwidth\\
			\hline
			$\sigma _S^2$, $\sigma _E^2$           & The power of AWGN at $S$ and $E$\\
			\hline
			${f_k}\left( n \right)$                & CPU frequency of $U_k$\\
			\hline
			$F_k^{\max }$, $F_S^{\max }$           & Peak CPU frequency of $U_k$ and $S$\\
			\hline
			${L_k}\left( n \right)$                & The remaining data of $U_k$ \\
			\hline
			${C_k}$, ${C_S}$                       & \makecell[l]{CPU cycles required for $U_k$ and $S$ to compute \\each bit task data}\\
			\hline
			${\varphi _k}$, ${\varphi _S}$         & The effective capacitance coefficient of $U_k$ and $S$\\
			\hline
			${E_S}\left( n \right)$                & The residual energy of $S$ \\
			\hline
			${w_1}$, ${w_2}$                       & \makecell[l]{The weighted factor of energy consumption and \\execution delay}\\
			\hline
			${c_E}$, ${c_T}$                       & \makecell[l]{Per unit cost of energy consumption and execution \\delay}\\
			\hline
			${Z_{\min }}$, ${Z_{\max }}$           & Flight altitude range of $S$\\
			\Xhline{1.2pt}
		\end{tabular}
	}
\end{table}

\section{System Model And Problem Formulation}
\label{sec:sysmod}

\begin{figure}[t]
	\centering
 	\includegraphics[width = 0.35 \textwidth]{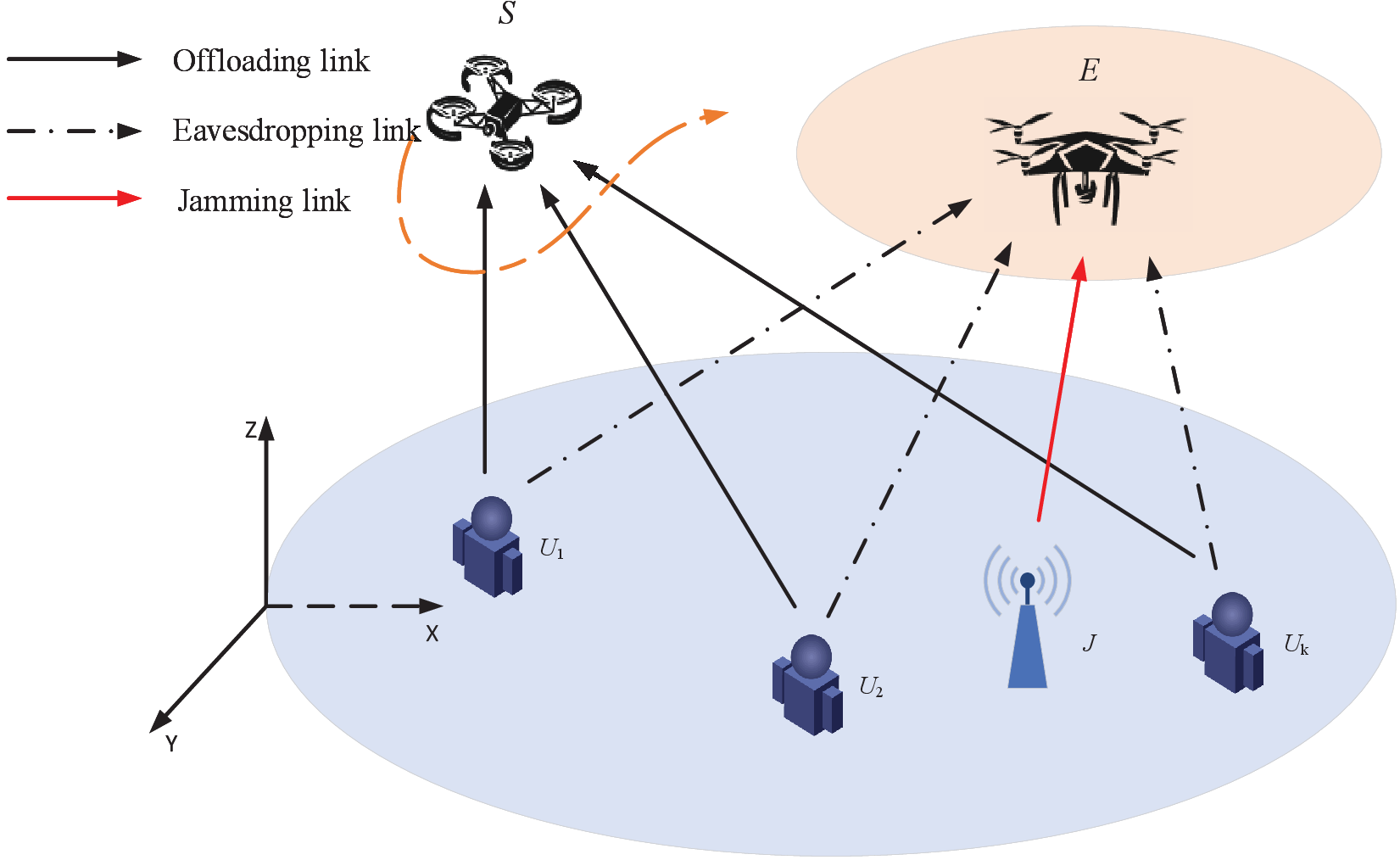}
	\caption{System model.}
	\label{figmodel}
\end{figure}

Fig. \ref{figmodel} shows the considered aerial MEC system, which consists of an aerial MEC server $\left( S \right)$, $K$ GUs ${U_k}$, denoted by the set $\mathcal{K} = \left\{ 1,2, \cdots, K \right\}$,
and an aerial malicious eavesdropper $\left( E \right)$.
A terrestrial friendly jammer $\left( J \right)$ is utilized to mitigate the eavesdropping capabilities of $E$ by transmitting artificial noise (AN).
All nodes are presumed to operate in half-duplex mode and are equipped with a single antenna.
The total flight time of $S$ is represented as $T$, which is equally divided into $N$ slots, and each slot with a duration of ${\delta _t} = \frac{T}{N}$.
Assume ${\delta _t}$ is small enough for treating the positions of $S$ as constant in each slot \cite{WuQ2018TCOM}.

A 3D Cartesian coordinate is utilized and the location of $S$ is expressed as
${{\mathbf{q}}_S}\left( n \right) = {\left[ {{x_S}\left( n \right),{y_S}\left( n \right),{z_S}\left( n \right)} \right]^T}$.
The coordinates of $J$ and ${U}_k$ at the $n$th slot are represented as
${{\mathbf{w}}_J} = {\left[ {{x_J},{y_J}} \right]^T}$ and
${{\mathbf{w}}_{{U_k}}} = {\left[ {{x_k},{y_k}} \right]^T}$, respectively.
{It is assumed that $E$ {works} at a certain altitude}\footnote{
	The aerial eavesdropper is assumed to be a malicious eavesdropper that passes through certain areas that contain sensitive signals. In particular, the aerial eavesdroppers could be disguised as a civilian aircraft and work at a fixed altitude. Considering the more practical scenario of $E$ working in a 3D trajectory is an interesting work and will be part of future work.} of ${z_E}$ and its exact location is unavailable for $S$ and $J$ {\cite{LeiH2022IOT}}.
Like \cite{ZhouY2020TCOM, LuW2022TCOM}, $E$ lies in a circle with center ${{\mathbf{\tilde q}}_E} = \left[
 {{{\tilde x}_E},{{\tilde y}_E}} \right] ^T$ and radius $r_E$, which satisfies $\left\| {{{\mathbf{q}}_E}\left( n \right) - {{{\mathbf{\tilde q}}}_E}} \right\| \le {r_E}$,
where
${{\mathbf{q}}_E}\left( n \right) = \left[ {{{x}_E}\left( n \right),{{y}_E}\left( n \right)} \right] ^T$ denotes the exact location of $E$.

\subsection{Secure Communication Model}

The ground-to-air (G2A) link between ${U}_k$ and $u$ $\left( {u \in \left\{ {S,E} \right\}} \right)$ is assumed to be a probabilistic LoS link and the probability is expressed as \cite{HouraniA2014WCL} \footnote{
	In Ref. \cite{LeiH2024PLOS}, the average rate with different 3D trajectories was compared, and the results demonstrated that there is an optimal 3D trajectory to maximize the average rate, which is one of the motivations of this work. 
}
\begin{align}
	&P_{k,u}^{{\mathrm{LoS}}}\left( n \right) \nonumber\\
	&\;\;= \frac{1}{{1 + {\eta _a}\exp \left( { - {\eta _b}\left( {\frac{\pi }{{180}}\arcsin \left( {\frac{{{z_u}\left( n \right)}}{{{d_{k,u}}\left( n \right)}}} \right)} - {\eta _a} \right) }\right)}},
	\label{LoSNP}
\end{align}
where
${d_{k,u}}\left( n \right) = \sqrt {{z_u}{{\left( n \right)}^2} + {{\left\| {\left[ {{x_u}\left( n \right),{y_u}\left( n \right)} \right]^T - {\mathbf{w}}_{{U_k}}} \right\|}^2}}$,
${\eta _a}$ and ${{\eta _b}}$ are environment parameters,
and
${z_u}\left( n \right)$ denotes the altitude of $u$.
The average path loss between ${U}_k$ and $u$ in dB is expressed as \cite{HouraniA2014WCL}
\begin{align}
	{L_{k,u}}\left( n \right) = P_{k,u}^{\mathrm {LoS}}\left( n \right)L_{k,u}^{\mathrm {LoS}}\left( n \right) + P_{k,u}^{\mathrm {NLoS}}\left( n \right)L_{k,u}^{\mathrm {NLoS}}\left( n \right),
	\label{Avepthloss}
\end{align}
{where
$P_{k,u}^{{\mathrm{NLoS}}}\left( n \right) = 1 - P_{k,u}^{{\mathrm{LoS}}}\left( n \right)$,
$L_{k,u}^{\mathrm {LoS}}\left( n \right) = L_{k,u}^{\mathrm {FS}}\left( n \right) + {\eta _{\mathrm {LoS}}}$ and
$L_{k,u}^{\mathrm {NLoS}}\left( n \right) = L_{k,u}^{\mathrm {FS}}\left( n \right) + {\eta _{{\mathrm{NLoS}}}}$ represent the pathloss of LoS and NLoS links in dB between ${U}_k$ and $u$, respectively,
$L_{k,u}^{{\mathrm{FS}}}\left( n \right) = 20{\log _{10}}{d_{k,u}}\left( n \right) + 20{\log _{10}}\left( {\frac{{4\pi {f_c}}}{c}} \right)$ is free space pathloss,
${f_c}$ is the carrier frequency,
${c}$ denotes the velocity of light,
${\eta _{\mathrm{LoS}}}$ and ${\eta _{\mathrm{NLoS}}}$ represent the excessive pathloss for LoS and NLoS links, respectively.}
The channel gain between ${U}_k$ and $u$ at the $n$th slot is expressed as
\begin{align}
	{h_{k,u}}\left( n \right) = {10^{- \frac{{ {L_{k,u}}\left( n \right)}}{{10}}}}.
	\label{ChannelGain}
\end{align}

{
	NOMA-based UAV-MEC systems not only can attain higher spectral efficiency and ensure the fairness of user offloading through the power allocation mechanism but also effectively can reduce the delay and energy consumption of user offloading \cite{ZhongR2022TWC}. Thus, it is assumed that users offload data to $S$ with NOMA technology. 
In particular, $S$ utilizes {channel state information (CSI)-based} successive interference cancellation (SIC) technology to decode user information in descending order of channel gain {\cite{ShiZ2022TCOM}}-\cite{LeiH2023IoTNOMA}. }
The signal-to-interference-plus-noise-ratio (SINR) of the signal from $U_k$ at $S$ is expressed as
\begin{align}
		{\gamma _{k,S}}\left( n \right) &= \frac{{{h_{k,S}}\left( n \right){p_k}\left( n \right)}}{{{h_{J,S}}\left( n \right){P_J} + \sum\limits_{l \ne k,l \in {\cal K}} {{\lambda _{k,l}}\left( n \right){h_{l,S}}\left( n \right){p_l}\left( n \right)}  + \sigma _S^2}} \nonumber \\
		&\mathop  = \limits^{\left( a \right)} \frac{{{h_{k,S}}\left( n \right){p_k}\left( n \right)}}{{\sum\limits_{l \ne k,l \in {\cal K}} {{\lambda _{k,l}}\left( n \right){h_{l,S}}\left( n \right){p_l}\left( n \right)}  + \sigma _S^2}},\forall k,n,
		\label{SNRkS}
\end{align}
{
	where
${h_{i,S}}\left( n \right)$ is the channel gain between $U_i$ and $S$,
$i \in \left\{ {k,l} \right\}$,
${p_i}\left( n \right)$ denotes the transmit power of $U_i$,
$\sigma_S ^2$ represents the additive white Gaussian noise (AWGN) power at $S$,
step $(a)$ holds if the friendly jammer is considered \footnote{{
	It is assumed that the AN signals emitted by $J$ conform to a Gaussian pseudo-random sequence since both $S$ and $J$ belong to a legitimate network. 
	Specifically, $ S$ has prior knowledge of the AN sent by $J$ and the AN signals can be ideally removed from the received signals at $S$. Because $E$ does not belong to the legitimate network, it is not aware of the existence of $J$ and treats all signals eavesdropped as the user's signals. Therefore, the AN signal sent by $J$ will interfere with $E$ \cite{XuY2021TCOM}, \cite{DingY2023TVT}, \cite{LuW2022TCOM}, \cite{LvL2019TIFS}, \cite{LeiH2023IoT}.
}},
}
and
${\lambda _{k,l}}\left( n \right)$ indicates the relationship between the channel gains of $U_k$-$S$ and $U_l$-$S$ in the $n$th slot, which is defined as \cite{XuY2021TCOM, LuW2022TCOM}
\begin{align}
	{\lambda _{k,l}}\left( n \right) = \left\{ {\begin{array}{*{20}{c}}
			{1,}&{{L_{k,S}}\left( n \right) < {L_{l,S}}\left( n \right),}  \\
			{0,}&{{L_{k,S}}\left( n \right) \ge {L_{l,S}}\left( n \right).}
	\end{array}} \right.
	\label{lamdakl}
\end{align}
{Specifically, ${\lambda _{k,l}}\left( n \right) = 1$ signifies that $U_k$-$S$ link has higher channel quality than $U_l$-$S$ link, and ${\lambda _{k,l}}\left( n \right) + {\lambda _{l,k}}\left( n \right) = 1$.}
Therefore, the instantaneous offloading rate from $U_k$ at $S$ is written as
\begin{align}
	{R_{k,S}}\left( n \right) = B{\log _2}\left( {1 + {\gamma_{k,S}}\left( n \right)} \right),
\end{align}
where $B$ is the channel bandwidth.

Similarly, the instantaneous offloading rate from $U_k$ at $E$ is expressed as
\begin{align}
	{R_{k,E}}\left( n \right) = B{\log _2}\left( {1 + {\gamma_{k,E}}\left( n \right)} \right),
\end{align}
where
$	{\gamma_{k,E}}\left( n \right) = \frac{{{h_{k,E}}\left( n \right){p_k}\left( n \right)}}{{{h_{J,E}}\left( n \right){P_J} + \sum\limits_{z \in {\mathcal{K}_k}} {{h_{z,E}}\left( n \right){p_z}\left( n \right)}  + \sigma_E ^2}}$,
${\mathcal{K}_k} = \left\{ {U_z| {{L_{z,E}}\left( n \right) \ge {L_{k,E}}\left( n \right)}} \right\}$ denotes the set whose channel gain is worse than the channel gain between ${U}_k$ and $E$,
{${h_{z,E}}\left( n \right)$ is the channel gain between $U_z$ and $E$,}
${h_{J,E}}\left( n \right)$ is the channel gain between $J$ and $E$,
${P_J}$ is the transmit power of $J$,
and $\sigma_E ^2$ represents the AWGN power at $E$.

Due to the uncertainty in the position of $E$, the exact value of ${R_{k,E}}\left( n \right)$ cannot be obtained \cite{ZhouY2020TCOM, LuW2022TCOM}.
To facilitate the analysis, we consider the eavesdropping rate $R_{k,E}^{{\mathrm{ub}}}\left( n \right)$ in the worst-case security situation.
By using the triangle inequality to scale the distance, the lower bound of the Euclidean distance between ${U}_k$ and $E$ is expressed as
\begin{align}
		{d_{k,E}} &= \sqrt {z_E^2 + {{\left( {\left\| {{{\mathbf{w}}_{{U_k}}}\left( n \right) - {{\mathbf{q}}_E}\left( n \right)} \right\|} \right)}^2}} \nonumber \\
		&\ge \sqrt {z_E^2 + {{\left( {\left\| {{{\mathbf{w}}_{{U_k}}}\left( n \right) - {{{\mathbf{\tilde q}}}_E}} \right\| - \left\| {{{{\mathbf{\tilde q}}}_E} - {{\mathbf{q}}_E}\left( n \right)} \right\|} \right)}^2}} \nonumber\\
		&\ge \sqrt {z_E^2 + {{\left( {\left\| {{{\mathbf{w}}_{{U_k}}}\left( n \right) - {{{\mathbf{\tilde q}}}_E}} \right\| - {r_E}} \right)}^2}} \buildrel \Delta \over =  d_{k,E}^{{\mathrm{lb}}}.
\end{align}
Similarly, the upper bound of the distance between $J$, $U_z$ $(z \in {\mathcal{K}_k})$ and $E$ are expressed as
\begin{align}
		{d_{J,E}} &= \sqrt {z_E^2 + {{\left( {\left\| {{{\mathbf{w}}_J} - {{\mathbf{q}}_E}\left( n \right)} \right\|} \right)}^2}} \nonumber\\
		&\le \sqrt {z_E^2 + {{\left( {\left\| {{{\mathbf{w}}_J} - {{{\mathbf{\tilde q}}}_E}} \right\| + \left\| {{{{\mathbf{\tilde q}}}_E} - {{\mathbf{q}}_E}\left( n \right)} \right\|} \right)}^2}} \nonumber\\
		&\le \sqrt {z_E^2 + {{\left( {\left\| {{{\mathbf{w}}_J} - {{{\mathbf{\tilde q}}}_E}} \right\| + {r_E}} \right)}^2}} \buildrel \Delta \over = d_{J,E}^{{\mathrm{ub}}}
\end{align}
and
\begin{align}
		{d_{z,E}} &= \sqrt {z_E^2 + {{\left( {\left\| {{{\mathbf{w}}_{{U_z}}}\left( n \right) - {{\mathbf{q}}_E}\left( n \right)} \right\|} \right)}^2}} \nonumber\\
		&\le \sqrt {z_E^2 + {{\left( {\left\| {{{\mathbf{w}}_{{U_z}}}\left( n \right) - {{{\mathbf{\tilde q}}}_E}} \right\| + \left\| {{{{\mathbf{\tilde q}}}_E} - {{\mathbf{q}}_E}\left( n \right)} \right\|} \right)}^2}} \nonumber\\
		&\le \sqrt {z_E^2 + {{\left( {\left\| {{{\mathbf{w}}_{{U_z}}}\left( n \right) - {{{\mathbf{\tilde q}}}_E}} \right\| + {r_E}} \right)}^2}} \buildrel \Delta \over= d_{z,E}^{{\mathrm{ub}}},
\end{align}
respectively.
Then we obtain 
\begin{align}
	&R_{k,E}^{{\mathrm{ub}}}\left( n \right) = B{\log _2}\left( {1 + \gamma _{_{k,E}}^{{\mathrm{ub}}}\left( n \right)} \right) \nonumber\\
	&= B{\log _2}\left( {1 + \frac{{h_{k,E}^{{\mathrm{ub}}}\left( n \right){p_k}\left( n \right)}}{{h_{J,E}^{{\mathrm{lb}}}\left( n \right){P_J} + \sum\limits_{z \in {\mathcal{K}_k}} {h_{z,E}^{{\mathrm{lb}}}\left( n \right){p_z}\left( n \right)}  + \sigma _E^2}}} \right),
\end{align}
where 
the superscript `ub' and `lb' signify the upper and lower bounds,
$h_{k,E}^{{\mathrm{ub}}}$, $h_{J,E}^{{\mathrm{lb}}}$, and $h_{z,E}^{{\mathrm{lb}}}$ are obtined by substituting $d_{k,E}^{{\mathrm{lb}}}$, $d_{J,E}^{{\mathrm{ub}}}$, and $d_{z,E}^{{\mathrm{ub}}}$ into (\ref{LoSNP}), (\ref{Avepthloss}), and (\ref{ChannelGain}), respectively.

The worst instantaneous secrecy offloading rate of $U_k$ in $n$th slot is expressed as
\begin{align}
	{R_k^{\sec }\left( n \right)} = {\left[ {{R_{k,S}}\left( n \right) - R_{k,E}^{{\mathrm{ub}}}\left( n \right)} \right]^ + },
\end{align}
where ${\left[ x \right]^ + } = \max \left\{ {x,0} \right\}$.
To prevent $E$ from eavesdropping, the following condition should be met \cite{LeiH2023IoT}, \cite{LiuS2022TWC}
\begin{align}
	{R_k^{\sec }\left( n \right)} \ge R_{\min }^{\sec },\forall k,n,
	\label{ratesec}
\end{align}
where
$R_{\min }^{\sec}$ denotes the threshold secure offloading rate.

\subsection{Mobile Edge Computing Model}

It is assumed that $U_k$ has the computationally-intensive and latency-sensitive task $L_k$ which needs to be completed within the flying time $T$, where ${L_k}$ represents the total amount of task data.
The computing tasks can not be finished locally due to the users' limited computing resources and battery capacity.
To solve this problem, users can offload part of the computational tasks to the aerial edge server for processing.
{Note that the computational tasks are bitwise independent and can be randomly separated to enable parallel computation between local execution and task offloading \cite{LiuB2022TVT}, \cite{JiJ2021IOT}.}

\emph{1) Local Computation:}
To fully utilize the power of local computing, $U_k$ and $S$ always adopt a dynamic voltage and frequency scaling technology \cite{XuY2021TWC}.
Therefore, the CPU frequency of $U_k$ is adjusted to reduce the energy consumed to perform local computation.
The amount of locally computed data and energy consumption for $U_k$ in the $n$th slot is given by
\begin{subequations}
	\begin{align}
		L_k^{\mathrm {loc}}\left( n \right) &= \frac{{{\delta _t}{f_k}\left( n \right)}}{{{C_k}}},\forall k,n,  \label{L_local}\\
		E_k^{\mathrm {loc}}\left( n \right) &= {\delta _t}{\varphi _k}f_k^3\left( n \right),\forall k,n, \label{E_local}
	\end{align}
\end{subequations}
where
${C_k}$ denotes the CPU cycles required to compute 1-bit task,
${f_k}\left( n \right)$ is the CPU frequency of $U_k$,
and
${\varphi _k}$ is the effective capacitance of $U_k$, depending on the $U_k$'s chip architecture \cite{SunC2021TWC}.

\emph{2) Computation Offloading:}
The amount of offloaded data from $U_k$ and the energy consumption for transmission at $U_k$ is denoted as
\begin{subequations}
	\begin{align}
		L_k^{{\mathrm{off}}}\left( n \right) &= {\delta _t}{R_k^{\sec }\left( n \right)},\forall k,n,  \label{Lkoff'}\\
		E_k^{\mathrm {off}}\left( n \right) &= {p_k}\left( n \right){\delta _t},\forall k,n. \label{Ekoff}
	\end{align}
\end{subequations}

The CPU frequency and energy to process $U_k$'s offloading data at $S$ is expressed as \cite{XuY2021TCOM}, \cite{LuW2022TCOM}
\begin{subequations}
	\begin{align}
		{f_{Sk}}\left( n \right) &= \frac{{L_k^{\mathrm {off}}\left( n \right){C_S}}}{{{\delta _t}}},\forall k,n,  \label{fsk}\\
		E_{S,k}^{\mathrm {com}}\left( n \right) &= {\delta _t}{\varphi _S}f_{Sk}^3\left( n \right),\forall k,n, \label{Eskcom}
	\end{align}
\end{subequations}
where
${C_S}$ and ${\varphi _S}$ signify the CPU cycles required by $S$ to compute 1-bit of data
and
the effective capacitance of $S$, respectively.
{At each slot, the sum of the amount of offloaded data must be less than that $S$ can deal with, which is expressed as}
\begin{align}
	{\sum\limits_{k = 1}^K {{f_{Sk}}\left( n \right)} \le F_S^{\max },\forall n,}
	\label{Lkoff}
\end{align}
{where $F_S^{\max }$ is the maximum computational frequency of $S$.}

The consumed propulsion energy of $S$ in $n$th slot is expressed as \cite{ZengY2019TWC}
\begin{align}
	E_S^{{\mathrm{fly}}}\left( n \right) = {P_v}\left( n \right){\delta _t},\forall n,
\end{align}
where
${P_v}\left( n \right) = {P_i}{\left( {\sqrt {1 + \frac{{{v^4}\left( n \right)}}{{4v_0^4}}}  - \frac{{{v^2}\left( n \right)}}{{2v_0^2}}} \right)^{\frac{1}{2}}} + \frac{1}{2}{d_0}\rho sA{v^3}\left( n \right) + {P_0}\left( {1 + \frac{{3{v^2}\left( n \right)}}{{U_{tip}^2}}} \right) $
is the propulsion power consumption of rotary-wing UAV $S$, 
where 
${P_0}$ and ${P_i}$ denote the blade profile and the induced power in hover state,
${U_{tip}}$ signifies the tip velocity of the rotor blade,
${v_0}$ denotes the mean rotor induced speed in hover,
$d_0$, $\rho$, $s$ and $A$ represent the drag ratio of the fuselage, the air density, solidity of the rotor and rotor disc area, respectively.

Based on \cite{YangZ2020TVT}, we only take into account the energy consumption of the propulsion and computation for executing user offloading tasks, while disregarding the energy consumption related to communication, which represents merely a small fraction of the UAV's total energy usage.
Thus, the residual energy of $S$ in $n$th slot is expressed as
\begin{align}
	{E_S}\left( n \right) = E_S^{\max } - \sum\limits_{i = 0}^{n - 1} {\left( {\sum\limits_{k = 1}^K {E_{S,k}^{{\mathrm{com}}}\left( i \right)}  + E_S^{{\mathrm{fly}}}\left( i \right)} \right)} ,
\end{align}
where $E_S^{\max }$ denotes the battery capacity of $S$.
{To ensure that all $U_k$'s tasks can be finished before $S$ runs out of the given energy, the following condition must be satisfied
\begin{align}
	{E_S}\left( N \right) \ge E_S^{\min },
	\label{ESN}
\end{align}
where 
${E_S^{\min}}$ denotes the minimum energy that ensure $S$ can fly directly back to the specified final position after the completion of the mission.}

{\emph{3) Computation Cost:}}
Similar to \cite{WangL2021TMC, ZhaoR2021TGCN, LuW2022TNSE}, the delay and energy consumption in returning processed results are neglected because of the substantial difference in transmitting power between $S$ and $U_k$ and the considerably smaller data volume generated by task calculation results than that caused by $U_k$. 
{When the computational tasks of all the GUs are completed, the energy consumption of the considered system is obtained as}
\begin{align}
	{{E_c} = \sum\limits_{n = 1}^N {\sum\limits_{k = 1}^K {\left( {E_k^{{\mathrm{loc}}}\left( n \right) + E_k^{{\mathrm{off}}}\left( n \right)} \right)}}.}
\end{align}

${T_k}\left( n \right)$ is defined as the state of $U_k$'s remaining unprocessed data in the $n$th slot, which is given as
\begin{align}
	{T_k}\left( n \right) = \left\{ {\begin{array}{*{20}{l}}
		{1,}&{{\mathrm{if}}\,{L_k}\left( n \right) > 0,}\\
		{0,}&{{\mathrm{otherwise}},}
		\end{array}} \right.
\end{align}
where
${L_k}\left( n \right)$ denotes the remaining data of $U_k$ in the $n$th slot, which is modeled as
\begin{align}
	{L_k}\left( n \right) = {\left[ {{L_k}\left( n - 1 \right) - \left( {L_k^{{\mathrm{loc}}}\left( n - 1 \right) + L_k^{{\mathrm{off}}}\left( n - 1 \right)} \right)} \right]^ + }.
	\label{user data remain}
\end{align}
Note that ${L_k}\left[ 0 \right] = {L_k}$ is satisfied.
Then the delay of $U_k$ is ${T_k} = {\delta _t}\sum\limits_{n = 1}^{N_k} {{T_k}\left( n \right)} $, where $N_k$ is the total slots for $U_k$ to process $L_k$. The latency of all users in the system is
\begin{align}
	{T_c} = \sum\limits_{k = 1}^K {{T_k}}.
\end{align}

Finally, the average computation cost, the weighted energy consumption and execution delay, is defined as \cite{LuW2022TCOM}
\begin{align}
	{U_c} = \frac{1}{K}\left( {{\omega _1}{c_E}{E_c} + {\omega _2}{c_T}{T_c}} \right),
	\label{AverCost}
\end{align}
where ${c_E}$ and ${c_T}$ denote the unit costs for energy consumption and time delay, respectively, 
${\omega _1} \in \left[ {0,1} \right]$ and ${\omega _2} \in \left[ {0,1} \right]$ are the weighting factors satisfying ${\omega _1} + {\omega _2} = 1$,
which determine the priority or importance on energy consumption and time delay.

\subsection{Problem Formulation}

In this work, the average computation cost of the considered system is minimized by jointly designing the $S$'s trajectory, $U_k$'s transmit power, and CPU frequency.
Let ${\mathbf{Q}} = \left\{ {{{\mathbf{q}}_S}\left( n \right), \forall n} \right\}$, ${\mathbf{P}} = \left\{ {{p_k}\left( n \right),\forall k,n} \right\}$, and ${\mathbf{F}} = \left\{ {{f_k}\left( n \right),\forall k,n} \right\}$.
The optimization problem is expressed as 
\begin{subequations}
	\label{optimizationproblem}
	\begin{align}
	\mathcal{P}_{1} :\; &\mathop {\mathop {\min }\limits_{{\mathbf{Q}},{\mathbf{P}},{\mathbf{F}}} {U_c}}  \nonumber \\
	{\mathrm{s.t.}}\;
	&{\left\| {{{\mathbf{q}}_S}\left( {n + 1} \right) - {{\mathbf{q}}_S}\left( n \right)} \right\| \le V_S^{\max }{\delta _t} , \forall n,}	\label{eq:1a}\\
	&{\left\| {{{\mathbf{q}}_S}\left( n \right) - {{\mathbf{q}}_E}\left( n \right)} \right\|} \ge d_{\min },\forall n,	\label{eq:1b}\\
	&{{{\mathbf{q}}_S}\left( 0 \right) = {{\mathbf{q}}_S^{\mathbf{I}}},}  \label{eq:1c}\\
	&{Z_{\min }} \le {z_S}\left( n \right) \le {Z_{\max }},\forall n,	\label{eq:1d}\\
	&0 \le {p_k}\left( n \right) \le {{P_k^{\max }}},\forall k,n,	\label{eq:1e}\\
	&0 \le {f_k}\left( n \right) \le F_k^{\max },\forall k,n,	\label{eq:1f}\\
	&{\sum\limits_{k = 1}^K {{L_k}\left( N \right)} = 0,}	\label{eq:1g}\\
	&(\textrm{\ref{lamdakl}}), (\textrm{\ref{ratesec}}), (\textrm{\ref{Lkoff}}), (\textrm{\ref{ESN}}),	\nonumber
	\end{align}
\end{subequations}
where
{
	${\mathbf{q}}_S^{\mathbf{I}}$ signifies the initial position of $S$, 
	$V_S^{\max}$ is the maximum speed of $S$, 
	$d_{\min }$ denotes the minimum safety distance, 
	${Z_{\min }}$ and ${Z_{\max }}$ denote the minimum and maximum altitude of $S$, respectively. 
	(\ref{eq:1a}) represents the maximum flight speed constraint of $S$, 
(\ref{eq:1b}) denotes the minimum collision avoidance distance between $S$ and $E$,  
(\ref{eq:1c}) is a constraint for the initial position of $S$,
(\ref{eq:1d}) is the flight altitude constraint of $S$,  
(\ref{eq:1e}) denotes the constraint of transmit power of $U_k$,
(\ref{eq:1f}) is the constraint of the CPU frequency of $U_k$ and $S$,
(\ref{lamdakl}) denotes the constraint for decoding ability of $U_k$, 
and 
(\ref{eq:1g}) is constraint for ensuring computation tasks are completed.}

The constraints (\ref{ratesec}) and (\ref{Lkoff}) are non-convex because of the coupling of two optimization variables, $\mathbf{Q}$ and $\mathbf{P}$.
And the constraint (\ref{ESN}) is also non-convex since the energy consumption model is involved. Furthermore, considering the vast solution space for decision-making and the dynamic environmental changes resulting from the mobility of UAV, optimization problem $\mathcal{P}_{1}$ is challenging to solve using traditional convex optimization methods.

\begin{figure*}[t]
	\centering
	\includegraphics[width = 0.7 \textwidth]{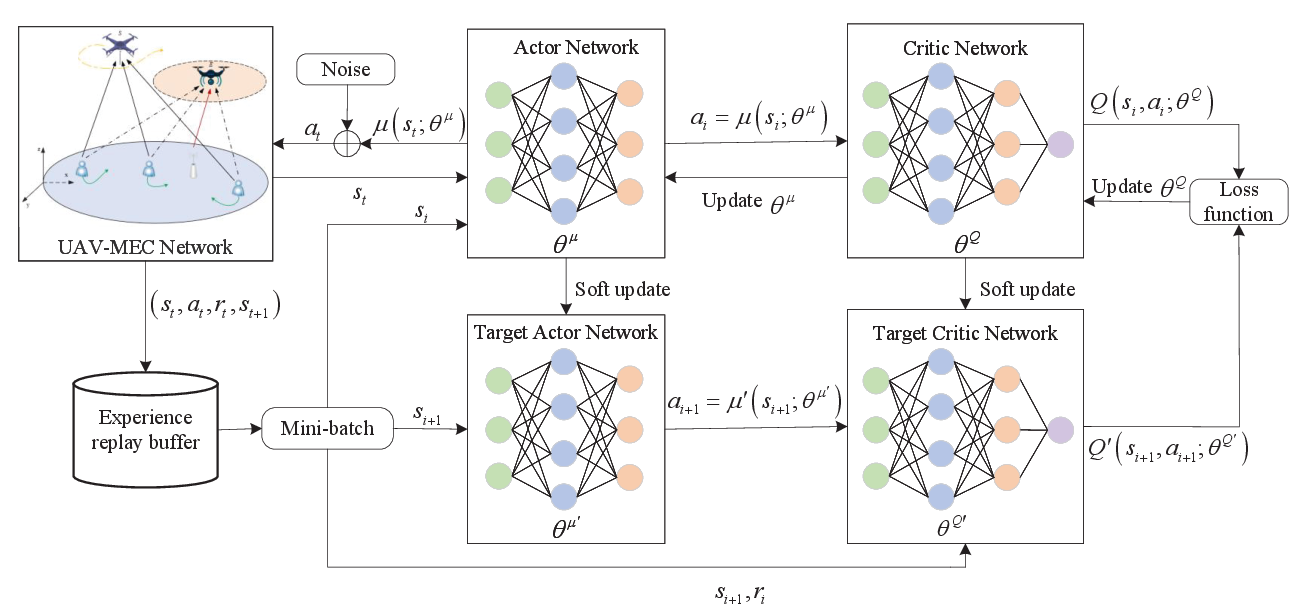}
	\caption{DDPG framework in the UAV-MEC network.}
	\label{DDPGmodel}
\end{figure*}

\section{DDPG-based Trajectory and Dynamic Resource Allocation Scheme}
\label{sec:DDPG}

DRL technology can adapt to dynamic environments and excel at tackling various non-convex and highly complex problems.
Therefore, to address the above mentioned issues, $\mathcal{P}_{1}$ is re-modeled as a single-agent extension of the Markov decision process (MDP), which is then solved by a DRL method as follows.

\subsection{MDP Formulation}

{Like \cite{DingR2020TWC, LeiH2024DDPG}, the spherical coordinate $\{ v, \theta, \varphi\}$ is utilized to describe the velocity and direction of $S$, where $v$ denotes the velocity,
$\theta $ represents the polar angle from the positive $z$-axis, constrained within the range $0 \le \theta  \le \pi $, while $\varphi $ denotes the azimuthal angle in the $xy$-plane, measured from the $x$-axis and constrained within the range $ 0  \le \varphi  \le 2\pi $.
Then, the velocity is expressed as
${\mathbf{v}}\left( n \right) = \left\{ {v\left( n \right),\theta\left( n \right),\varphi\left( n \right)} \right\}$.
Define the distance of $S$ within two consecutive slots as
$v\left( n \right){\delta _t} = \left\| {{{\mathbf{q}}_S}\left( n + 1 \right)- {{\mathbf{q}}_S}\left( n \right)} \right\|$, we have}
\begin{subequations}
	\begin{align}
		{x_S}\left( n + 1 \right) &= {x_S}\left( n \right) + {v\left( n \right){\delta _t}}\sin \theta\left( n \right) \cos \varphi\left( n \right), 	\label{x0}\\
		{y_S}\left( n + 1 \right) &= {y_S}\left( n \right) + {v\left( n \right){\delta _t}}\sin \theta\left( n \right) \sin \varphi\left( n \right),	\label{y0}\\
		{z_S}\left( n + 1 \right) &= {z_S}\left( n \right) + {v\left( n \right){\delta _t}}\cos \theta\left( n \right).		\label{z0}
	\end{align}
\end{subequations}
At each slot, $S$ determines its speed $v\left( n \right)$, flight direction $\theta \left( n \right)$ and $\varphi \left( n \right)$, the transmit power ${p_k}\left( n \right)$ and computational frequency ${f_k}\left( n \right)$ of each $U_k$ to obtain the minimum average computational cost for system users.
Since $S$'s movement will affect its environment, the users' total computational cost depends on the environment's current state and the UAV's actions.
Furthermore, the previous state and actions of the UAV can together result in a new stochastic state of the environment. 

As a result, $\mathcal{P}_{1}$ can be represented as a MDP, which is defined as a four-tuple
$\left\langle {\mathcal{S},\mathcal{A},\mathcal{P},\mathcal{R} }\right\rangle$, where $\mathcal{S}$, $\mathcal{A}$, $\mathcal{P}$, and $\mathcal{R}$ denote the state space, action space, state transition probability, and reward, respectively, which are presented as follows.

\emph{1) The state space $\left( {\mathcal{S}} \right)$:}
The state $s\left( n \right)$ is composed of ${\mathbf {q}_S}\left( n \right)$, ${E}_S\left( n \right)$, ${R_k^{\sec }\left( n \right)}$ and ${L_k}\left( n \right)$, that is,
\begin{align}
	s\left( n \right) = \left\{ {{{\mathbf{q}}_S}\left( n \right),{E}_S\left( n \right),{R_k^{\sec }\left( n \right)},{L_k}\left( n \right),\forall k} \right\}.
\end{align}

\emph{2) The action space $\left( {\mathcal{A}} \right)$:}
Since $S$ is required to determine its movements including the flight speed $v\left( n \right)$, the polar angle $\theta \left( n \right)$, the azimuthal angle $\varphi \left( n \right)$, the $U_k$'s transmit power ${p_k}\left( n \right)$ and computational frequency ${f_k}\left( n \right)$. The action space $a\left( n \right)$ is designed by
\begin{align}
	a\left( n \right) = \left\{ {v\left( n \right),\theta \left( n \right),\varphi \left( n \right),{p_k}\left( n \right),{f_k}\left( n \right),\forall k} \right\}.
\end{align}
The ranges of each element in $a\left( n \right)$ are expressed as, $v\left( n \right) \in \left[ {0,V_S^{\max }} \right]$, $\theta \left( n \right) \in \left[ {0,\pi } \right]$, $\varphi \left( n \right) \in \left[ { 0 ,2\pi } \right]$, ${p_k}\left( n \right) \in \left[ {0,{{P_k^{\max }}}} \right]$,
and
${f_k}\left( n \right) \in \left[ {0,F_k^{\max }} \right]$.

It can be observed that the state space $\mathcal{S}$ has ($4 + 2K$) dimensions, and the action space $\mathcal{A}$ is a continuous set with ($3 + 2K$) dimensions.
Moreover, with increasing $K$, the size of $\mathcal{S}$ and $\mathcal{A}$ exponentially increases.

\emph{3) The reward $\left( {\mathcal{R}} \right)$:}
According to the optimization problem (\ref{optimizationproblem}), the reward of this work is designed by comprising the following components:
the offloading data reward,
the satisfying constraints reward,
and
the computational reward, which are introduced as follows.
\begin{itemize}

\item The offloading data reward:
Since $U_k$ can reduce their own computational costs by offloading tasks to $S$, the reward $\left( {{r_{{\rm{off}}}}\left( n \right)} \right)$ is designed as
\begin{align}
	{r_{{\mathrm{off}}}}\left( n \right) = {\kappa _{\mathrm{f}}}{\delta _t}\sum\limits_{k = 1}^K {{\vartheta _{k}\left( n \right)}{R_k^{\sec }\left( n \right)}},
\end{align}
where ${\kappa _{\mathrm{f}}}$ denotes a positive constant utilized for the adjustment of ${r_{{\mathrm{off}}}}\left( n \right)$
and
$\vartheta$ is applied to control the reward gained from $U_k$, which is expressed as
\begin{align}
	{\vartheta _{k}\left( n \right)} = \left\{ {\begin{array}{*{20}{l}}
			{1,}&{{\mathrm{if}}\;{L_k}\left( n \right) > 0,}\\
			{0,}&{{\mathrm{otherwise.}}}
	\end{array}} \right.
	\label{}
\end{align}

\item {The satisfying constraints reward: 
To ensure the constraints of the optimization problem are satisfied, 
the satisfying constraints reward 
(${r_{\mathrm{p}}}\left( n \right)$) is designed as
\begin{align}
	{r_{\mathrm{p}}}\left( n \right) =  - {\lambda _{{\mathrm{ac}}}}\left( n \right){\kappa _{{\mathrm{ac}}}} - {\lambda _{{\mathrm{rc}}}}\left( n \right){\kappa _{{\mathrm{rc}}}} - {\kappa _{{\mathrm{er}}}}\left( n \right),
	\label{conrew}
\end{align}
where ${\kappa _{{\mathrm{ac}}}}$ and ${\kappa _{{\mathrm{rc}}}}$
denote the rewards related with the collision constraint (\ref{eq:1b})
and
$S$'s computing resource constraints (\ref{Lkoff}), respectively,
${\lambda _{{\mathrm{ac}}}}\left( n \right)$ and ${\lambda _{{\mathrm{rc}}}}\left( n \right)$ are binary coefficients,
${\lambda _{{\mathrm{ac}}}}\left( n \right) = 1$ denotes (\ref{eq:1b}) is not satisfied,
${\lambda _{{\mathrm{rc}}}}\left( n \right) = 1$ denotes (\ref{Lkoff}) is not met,
${\kappa _{{\mathrm{er}}}}\left( n \right)$ is a sparse reward to judge the data remaining of all $U_k$ when $S$'s energy has been exhausted and defined as
\begin{align}
	{\kappa _{{\mathrm{er}}}}\left( n \right) = \left\{ {\begin{array}{*{20}{c}}
			{0,}&{n = 1, \cdots ,N - 1,}\\
			{\zeta \sum\limits_{k = 1}^K {{L_k}\left( N \right)},}&{n = N,}
	\end{array}} \right.
	\label{eq:P1}
\end{align}
where $\zeta $ is a positive constant that is used to adjust ${\kappa _{{\mathrm{er}}}}\left( n \right)$.
A negative reward will be given if constraint (\ref{eq:1g}) is not met; otherwise, no reward will be provided.
It should be noted that all the terms in (\ref{conrew}) always be negative.
This is because negative rewards imply punishment when constraints are not met.
}

\end{itemize}

Thus, the reward function in this work is expressed as
\begin{align}
	r\left( n \right) =  {r_{{\mathrm{off}}}}\left( n \right) + {r_{\mathrm{p}}}\left( n \right) - {U_c}\left( n \right),
\end{align}
where
${U_c}\left( n \right)$
denotes the computation cost at the $n$th slot.

At each time step, $S$ transitions from $s\left( n \right)$ to $s\left( n +1 \right)$ based on $a\left( n \right)$ and $\mathcal{P}$, and then receives a reward $r\left( n \right)$.
The sequential transitions $\left( {s\left( n \right),a\left( n \right),r\left( n \right),s\left( n +1 \right)} \right)$ are stored as experiences in a replay buffer and randomly sampled to train neural networks (NNs).
Then $S$ continuously selects the optimal action through the trained NN to maximize the cumulative discount reward.

\subsection{DDPG-Based Trajectory and Dynamic Resource Allocation Scheme}

Considering the high-dimensional continuous $a\left( n \right)$ of the formulated problem, the DDPG-based TDRA algorithm is proposed to address the above proposed MDP. 
The aim is to enable the agent to learn an optimal policy that maximizes the cumulative reward $\mathcal{R}_n$ by jointly optimizing the policy and the value function networks, which is defined as \cite{SilverD2014CML}
\begin{align}
	\mathcal{R}_n = \sum\limits_{n = 0}^{N - 1} {{\beta ^n}{r\left( n \right)}},
\end{align}
where $\beta  \in \left[ {0,1} \right]$ represents the discount factor,  which determines the trade-off between current and future rewards.

As illustrated in Fig. \ref{DDPGmodel}, the DDPG-based TDRA algorithm is implemented with an actor-critic architecture containing an actor NN and a critic NN. The actor NN is inputted with the current environmental state ${s\left( n \right)}$ and then provides the corresponding action $a\left( n \right)$ to the NOMA-aided aerial MEC network based on the weight ${\theta ^\mu }$ of the actor NN and a stochastic noise $\mathcal{N}_n$, the instantaneous action $a\left( n \right)$ is given by \cite{SilverD2014CML}
\begin{align}
	{a\left( n \right)} = \mu \left( {{s\left( n \right)};{\theta ^\mu }} \right) + \mathcal{N}_n,
\end{align}
where $\mathcal{N}_n$ is exploratory random noise to increase action randomness and diversity and enable more effective state space exploration.

During training, $\mathcal{N}_n$ decreases gradually, allowing the actor NN to transition from exploration to exploitation, effectively leveraging learned knowledge and experience. It is worth noting that ${a\left( n \right)}$ is constrained within the range $\left[ {0,1} \right]$. Therefore, it will be clipped if the noise-added action value exceeds the desired range. Subsequently, $S$ executes the current action ${a\left( n \right)}$. The environment is transferred to the next state $s\left( {n + 1} \right)$ according to a certain state transition probability. At the same time, the environment returns an immediate reward ${r\left( n \right)}$ to the agent according to the quality of ${a\left( n \right)}$.

To stabilize the training process and break the correlation of input data, $S$ stores the current experience $\left( {s\left( n \right),a\left( n \right),r\left( n \right),s\left( n +1 \right)} \right)$ in the replay buffer $\mathcal{B}$ with size ${M_r}$ and randomly samples a mini-batch of ${M_b}$ experience samples from $\mathcal{B}$ to train the actor NN and the critic NN, to approximate the action function and the action-value function, respectively. 

Furthermore, the pair $\left( {s\left( n \right),a\left( n \right)} \right)$ will be fed into the critic NN to evaluate ${a\left( n \right)}$. The critic NN with weight ${\theta ^Q}$ will output an estimated Q-value $Q\left( {s\left( n \right),a\left( n \right);{\theta ^Q}} \right)$, where the Q-value is the expected long-term reward \cite{SilverD2014CML}.
To enhance training stability and facilitate convergence, the DDPG-based TDRA algorithm incorporates target actor NN and target critic NN with weights ${\theta ^{\mu'}}$ and ${\theta ^{Q'}}$, respectively.

Based on the principle of the policy gradient theorem \cite{SilverD2014CML}, the goal of the DDPG-based TDRA algorithm is to maximize the discounted cumulative reward $J\left( \mu  \right)$. 
The weights ${\theta ^\mu }$ of the actor NN are updated along the direction of the gradient that improves the action-value $Q\left( {s\left( n \right),a\left( n \right);{\theta ^Q}} \right)$, which is obtained as (\ref{policygradient}), shown at the top of next page.

\setcounter{equation}{36}
\begin{figure*}[ht]
	\begin{align}
			{\nabla _{{\theta ^\mu }}}J &\approx E\left[ {{\nabla _{{\theta ^\mu }}}Q\left( {s,a;{\theta ^Q}} \right){|_{s = {s\left( n \right)},a = \mu \left( {{s\left( n \right)};{\theta ^\mu }} \right)}}} \right] \nonumber\\
			&= E\left[ {{\nabla _a}Q\left( {s,a;{\theta ^Q}} \right){|_{s = {s\left( n \right)},a = \mu \left( {{s\left( n \right)}} \right)}}{\nabla _{{\theta ^\mu }}}\mu \left( {{s\left( n \right)};{\theta ^\mu }} \right){|_{s = {s\left( n \right)}}}} \right]
			\label{policygradient}
	\end{align}
	\hrulefill
\end{figure*}

The critic NN is updated to minimize the loss function $L\left( {{\theta ^Q}} \right)$, which is defined as
\begin{align}
	L\left( {{\theta ^Q}} \right) = {\left( {{y_i} - Q\left( {{s_i},{a_i};{\theta ^Q}} \right)} \right)^2},
	\label{loss}
\end{align}
where ${y_i}$ is the target Q-value output by the target critic NN, defined as
\begin{align}
	{y_i} = {r_i} + \beta Q\left( {{s_{i + 1}},{\mu'}\left( {{s_{i + 1}};{\theta ^{\mu'}}} \right);{\theta ^{Q'}}} \right).
\end{align}

In terms of the target NNs, they are replicas of the original networks but with a slower update frequency, which allows for a more stable training of the actor and critic NNs.
The update rule for soft update is given by \cite{SilverD2014CML}
\begin{subequations}
	\begin{align}		
		{\theta ^{Q'}} &\leftarrow \tau {\theta ^Q} + \left( {1 - \tau } \right){\theta ^{Q'}}, \label{target critic weight}\\
		{\theta ^{\mu'}} &\leftarrow \tau {\theta ^\mu } + \left( {1 - \tau } \right){\theta ^{\mu'}}, \label{target actor weight}
	\end{align}
\end{subequations}
where $0 < \tau  \ll 1$ denotes the updating rate.
In accordance with the above settings, a DDPG-based TDRA algorithm is presented in Algorithm \ref{alg:algorithm1}.

\begin{algorithm}[t]
	\caption{DDPG-based TDRA Algorithm for $\mathcal{P}_{1}$}
	\label{alg:algorithm1}
	Randomly initialize $\mu \left( {s;{\theta ^\mu }} \right)$ and $Q\left( {s,a;{\theta ^Q}} \right)$.
	Initialize the associated target NNs with ${\theta ^{\mu'}} \leftarrow {\theta ^\mu }$, ${\theta ^{Q'}} \leftarrow {\theta ^Q}$. \\
	Initialize the experience replay buffer $\mathcal{B}$ with size ${M_r}$. \\
	Initialize discount factor $\beta $, mini-batch size with ${M_b}$, soft update coefficient $\tau $. \\
	\For {{episode} = $1,2,...,{M_{{\mathrm{ep}}}}$}{
		Initialize $\mathcal{N}_n$, ${{\mathbf{q}}_S}\left( 0 \right)$, $E_S^{\max }$, ${{\mathbf{w}}_{{U_k}}}\left( 0 \right)$, and ${L_k}\left( 0 \right)$. \\
		\While{${L_k}\left( n \right) > 0$}
			{\textbf{Input} state $s\left( n \right)$ into actor NN and obtain action ${a\left( n \right)} = \mu \left( {{s\left( n \right)};{\theta ^\mu }} \right) + \mathcal{N}_n$. \\
			Execute action $a\left( n \right)$, and then receive reward $r\left( n \right)$ and observe the next state $s\left( {n + 1} \right)$. \\
			Store experience tuple $\left( {s\left( n \right),a\left( n \right),r\left( n \right),s\left( n +1 \right)} \right)$ into $\mathcal{B}$. \\
			\If{{\rm{Memory counter}} $>{M_r}$}
				{Remove previous experience from the front of $\mathcal{B}$.}
			Randomly sample a mini-batch of ${M_b}$ tuples $\left( {{s_i},{a_i},{r_i},{s_{i + 1}}} \right)$ from $\mathcal{B}$.\\
			Update the critic NN's weight ${\theta ^Q}$ by minimizing the critic loss function (\ref{loss}). \\
			Update the actor NN's weight ${\theta ^\mu }$ according to (\ref{policygradient}). \\
			Update the target critic NN's weight ${\theta ^{Q'}}$ according to (\ref{target critic weight}). \\
			Update the target actor NN's weight ${\theta ^{\mu'}}$ according to (\ref{target actor weight}). \\
			\If{${E_S}\left( N \right) \le E_S^{\min }$}
			{break.}
			}
}
\end{algorithm}

According to \cite{WangK2023TWC}, \cite{YangH2021TWC}, the computational complexity of the proposed DDPG-based TDRA algorithm depends on the size of the mini-batch and the NN, the number of neurons, training steps, and the training episodes, which is expressed as $\mathcal{O}\left( {{M_{{\textrm{ep}}}}N{M_b}\sum\nolimits_{i = 0}^{L - 1} {{Z_i}{Z_{i + 1}}} } \right)$, where $L$ denotes the number of network layers and ${Z_i}$ is the number of neurons in the $i$th layer.

\section{Performance Evaluation}
\label{sec:Evaluation}

{
	To demonstrate the performance of the proposed DDPG-based TDRA scheme, simulation results are presented in this section. The main simulation parameters are shown in TABLE \ref{table3} \cite{LuW2022TCOM, DingR2020TWC}. 
}

\begin{table}[t]
	\centering\caption{Simulation Parameters}
	{
	\label{table3}
	\begin{tabular}{c|c|c|c}
		\Xhline{1pt}
		\textbf{Parameters} & \textbf{Value} & \textbf{Parameters} & \textbf{Value}\\
		\Xhline{1pt}
		${V_S^{\max }}$ & $20$\;m/s & ${P_k^{\max }}$ & 0.1\;W  \\
		\hline
		$B$ & 1\;MHZ &  $L_k$ & 100\;Mbit  \\
		\hline
		$F_k^{\max}$ & 0.1\;GHZ  & $F_S^{\max}$ & 20\;GHZ \\
		\hline
		$\delta_t$ & 0.5\;s  & $\sigma _S^2, \sigma _E^2$ & --100\;dBm \\
		\hline
		$R_{\min }^{\sec}$ & 0.9\;Mbps  &	$C_k,C_S$ & 1000\;cycles/bit  \\
		\hline
		$\varphi_k,\varphi_S$ & $10^{-28}$ &  $E_S^{\max } - E_S^{\min }$ & 20000 J\\
		\hline
		$c_E$ & 1 & $c_T$ & 1 \\
		\hline
		$\eta_a$ & 12.08 & $\eta_b$ & 0.11 \\
		\hline
		$\eta_{\mathrm{Los}}$ & 1.6\;dB  & 	$\eta_{\mathrm{NLos}}$ & 23\;dB \\
		\hline
		${Z_{\min }}$ & $100$\;m  & ${Z_{\max }}$ & $150$\;m \\
		\hline
		${{\mathbf{w}}_J}$ & ${\left[ {300,250,0} \right]^T}$ & ${\mathbf{q}}_S^{\rm I}$ & ${\left[ {0,250,100} \right]^T}$ \\
		\hline
		${{\mathbf{\tilde q}}_E}$ & ${\left[ {290,150,100} \right]^T}$ & ${r_E}$ & $25$\;m \\
		\Xhline{1pt}
	\end{tabular}}
\end{table}

\subsection{Network Architecture}
The actor NN and critic NN employ six fully-connected neural networks, including six hidden layers with 64, 128, 256, 256, 128 and 64 neurons, respectively.
The activation function for all layers is the ReLU except for the output layer of the actor NN, which uses a sigmoid.
It is worth noting that the states input to NN include the $S$'s position ${\mathbf {q}_S}\left( n \right)$ and the residual energy ${E}_S\left( n \right)$, the $U_k$'s achievable security rate ${R_k^{\sec }\left( n \right)}$ and residual data ${L_k}\left( n \right)$ with dimensions of 3, 1, $K$ and $K$ respectively.
In our simulation, $K$ is set as 5, which leads to the UAV state dimensions having a relatively small proportion within the overall state dimensions. This may need to be more effective in affecting the training process of the NN.
However, ${\mathbf {q}_S}\left( n \right)$ and ${E}_S\left( n \right)$ are of great importance, and by adjusting the trajectory of $S$, it needs to be ensured that all the $U_k$'s task is processed efficiently before $S$ runs out of energy.
Thus, there exists the dimension imbalance issue,  and to address it, we expand ${\mathbf {q}_S}\left( n \right)$ from 3 to 8 dimensions and the ${E}_S\left( n \right)$ from 1 to 5 dimensions to ensure that each input state dimension at one level. After dimension expansion, the input layer of the actor NN has $2K+13$ neurons, and the critic NN has $4K+16$ neurons.
{The corresponding training and reward parameters are in TABLE \ref{table4}.}

\begin{table}[t]
 	\centering\caption{Training Parameters}
 	{
 	\label{table4}
 	\begin{tabular}{c|c}
 		\Xhline{1pt}\textbf{Hyperparameters} & \textbf{Value} \\
 		\Xhline{1pt}
 		Total episodes {${M_{{\mathrm{ep}}}}$}            & 1000  \\
 		\hline
 		Memory capacity $M_r$         & 10000  \\
 		\hline
 		Mini-batch size $M_b$         & 128  \\
 		\hline
 		Actor NN learning rate   & 0.0001 \\
 		\hline
 		Critic NN learning rate  & 0.0006 \\
 		\hline
 		Updating rate $\tau$          & 0.001  \\
 		\hline
 		Discount factor $\beta $      & 0.99 \\
 		\hline
 		Optimizer                     & Adam Optimizer \\
 		\hline
 		${\kappa _{\mathrm{f}}}$          & $2.5 \times {10^{ - 7}}$ \\
 		\hline
 		${\kappa _{{\mathrm{ac}}}}$       & 1 \\
 		\hline
        ${\kappa _{{\mathrm{rc}}}}$       & 10 \\
 		\hline
 		$\zeta $                      & $10^{ - 7}$ \\
 		\Xhline{1pt}
 	\end{tabular}}
\end{table}

{
	Due to the varying value ranges of different states, the direct input to the NN will lead to problems such as unstable training process, disappearing or exploding gradient, and difficult convergence of the model. Therefore, before inputting the state values into the NN training, each state's value needs to be normalized to $\left[ {0,1} \right]$ to accelerate the NN's training speed, which is implemented as follows: The state values $\left\{ {{x_S}\left( n \right),{y_S}\left( n \right),{z_S}\left( n \right),{E_S}\left( n \right),R_k^{\sec }\left( n \right),{L_k}\left( n \right)} \right\}$ are divided by $\left\{ {{X_{\max }},{Y_{\max }},{Z_{\max }},E_S^{\max },R_{\max }^{\sec },{L_0}} \right\}$, respectively, where $\left({X_{\max }},{Y_{\max }},{Z_{\max }}\right)$ is the operation range of the UAV, and ${R_{\max }^{\sec }}$ is the maximum secure offloading rate that the user can achieve.
}

{
	Since the output layer of the actor network uses a sigmoid activation function to scale the range of the action values to $\left[ {0,1} \right]$, it is necessary to scale each action value to the actual range of values. The specific implementation is as follows: the action values $\left\{ {v\left( n \right),\theta \left( n \right),\varphi \left( n \right),{p_k}\left( n \right),{f_k}\left( n \right)} \right\}$ are multiplied by $\left\{ {V_S^{\max },\pi ,2\pi ,P_k^{\max },F_k^{\max }} \right\}$, respectively. In this way, by multiplying the corresponding maximum values, the standardized action values can be converted into a range of values for practical applications to ensure that the UAV and the user can operate and compute as expected.
	}
\subsection{Simulation Results and Analysis}

{
	To evaluate the performance of the proposed scheme, a baseline scheme is utilized as the benchmark wherein all the users offload data to $S$ with TDMA technology. 
	Specifically, in the benchmark scheme, the time slot is divided equally among $K$ users, and the trajectory design, offloading strategy, and power allocation are re-implemented using the same hyperparameters as the proposed algorithm and the reward design.
	The TDMA-based scheme is chosen as the benchmark since it is a typical representative of the OMA scheme and was considered in many works, such as \cite{XuY2021TCOM}, \cite{MaoW2023TWC}, \cite{LuW2021TII}, \cite{ZhaoR2021TGCN}. The results of the TDMA-based scheme in this work can also provide necessary guidance and benchmarks for other OMA schemes. 
}
The instantaneous secrecy offloading rate of the benchmark scheme is expressed as 
\begin{align}
	&R_k^{{\mathrm{TDMA}},{\mathrm{sec}}}\left( n \right) = \frac{B}{K}\left[ {{{\log }_2}\left( {1 + \frac{{{h_{k,S}}\left( n \right){p_k}\left( n \right)}}{{\sigma _S^2}}} \right) } \right. \nonumber\\
	& \;\;\;\;\;\;\;\;\;\;\;\;\;\;\;\;\;\;\;\;\;  - \left. {{{\log }_2}\left( {1 + \frac{{h_{k,E}^{{\rm{ub}}}\left( n \right){p_k}\left( n \right)}}{{h_{J,E}^{{\rm{lb}}}\left( n \right){P_J} + \sigma _E^2}}} \right)} \right].
	\label{tdmasc}
\end{align}

\begin{figure}[t]	
	\centering	
	\subfigure[Accumulated reward.] {
		\label{fig03a}
		\includegraphics[width = 0.35 \textwidth]{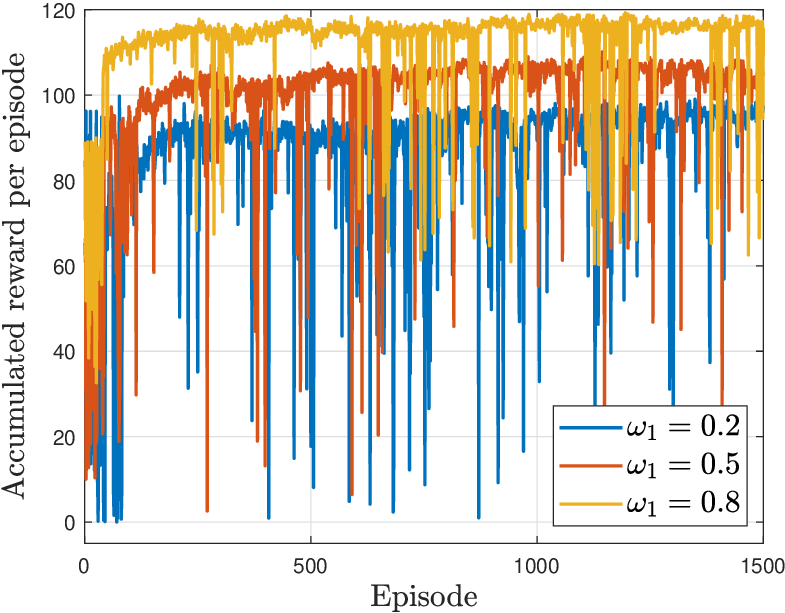}}
	\subfigure[Average cost.] {
		\label{fig03b}
		\includegraphics[width = 0.35 \textwidth]{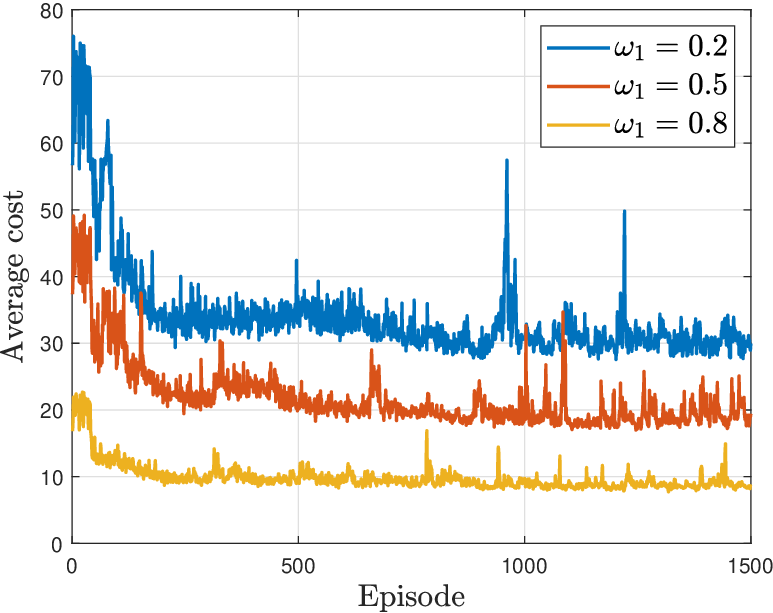}}
	\caption{Accumulated reward and average cost with varying ${\omega _1}$.}
	\label{3Converge}	
\end{figure}

Fig. \ref{3Converge} illustrates the accumulative reward and average cost of the proposed DDPG-based TDRA algorithm with varying ${\omega _1}$.
One can find that the reward curves converge as training round increases.
The accumulative reward of the proposed offloading scheme outperforms that of the benchmark scheme.
This is because the proposed scheme allows multiple users to transmit data simultaneously, so the amount of data offloaded for the proposed scheme is larger than that of the benchmark scheme, which will obtain a higher offloading data reward.
Furthermore, the accumulative reward of both schemes increases as $\omega_1$ increases.
This is attributed to the increasing energy cost in the average cost as $\omega_1$ grows, resulting in the reward function primarily descending toward reducing user energy costs.
Consequently, the users opt to offload more data to $S$, thus lowering their energy costs.
Fig. \ref{fig03b} demonstrates the average cost with varying ${\omega _1}$.
The average cost converges stably as training round increases.
Specifically, during the initial random exploration phase, there is almost no reduction in the cost due to the higher probability of random actions, and the training will start once the replay memory buffer is filled.
Moreover, the cost of the benchmark scheme begins to degrade earlier than the proposed scheme.
This is because, under the same experience replay buffer size, the benchmark scheme fills the experience pool and starts training faster than NOMA because it requires longer processing time when handling the same user data, resulting in more experiences generated in one episode for the benchmark.

\begin{figure}[t]
	\centering	
	\subfigure[${\omega _1}=0.2$] {
		\label{cost1}
		\includegraphics[width = 0.35 \textwidth]{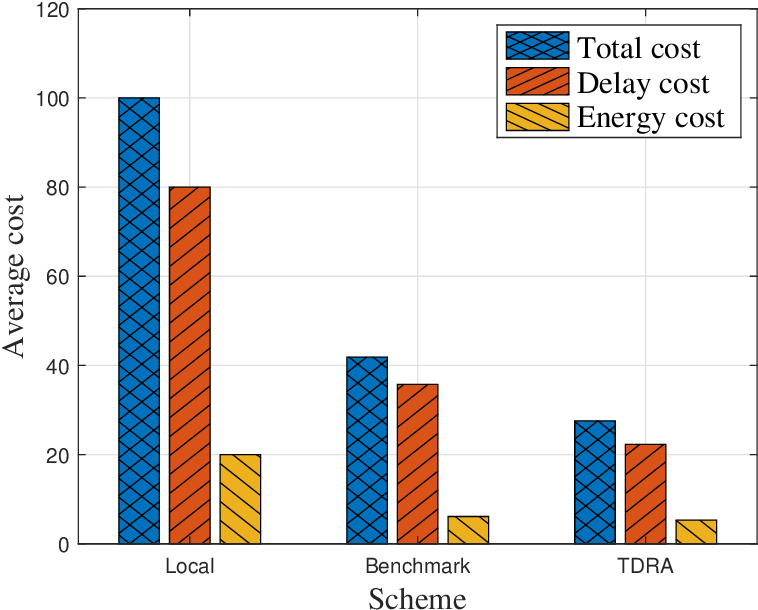}}
	\subfigure[${\omega _1}=0.5$] {
		\label{cost2}
		\includegraphics[width = 0.35 \textwidth]{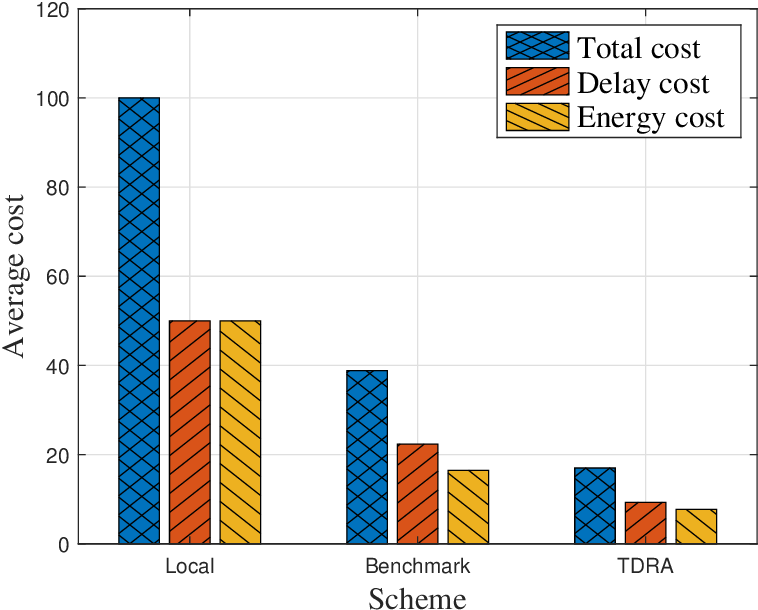}}
	\subfigure[${\omega _1}=0.8$] {
		\label{cost3}
		\includegraphics[width = 0.35 \textwidth]{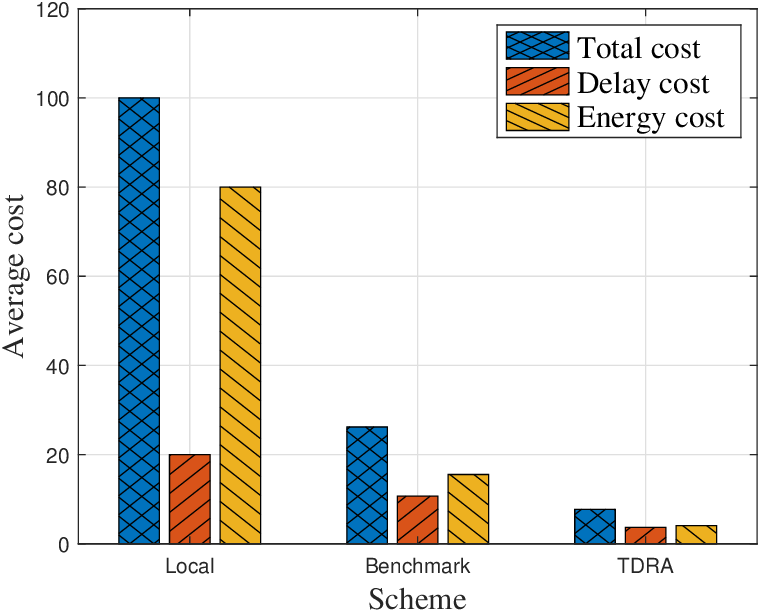}}
	\caption{The trend of energy consumption and delay costs with varying ${\omega _1}$.}
	\label{4Cost}
\end{figure}

Fig. \ref{4Cost} presents the trend of energy consumption and delay costs with varying ${\omega _1}$ wherein 
`Local' denotes that all the $U_k$ perform all computational tasks locally with the maximum computation frequency.
It can be seen that under the same weight, the cost of local scheme is the highest, followed by the benchmark, while the proposed scheme obtains the lowest average cost.
This is because fully local computation processes all user data locally without the assistance of MEC servers, resulting in the highest cost while the proposed and benchmark schemes partially offload user data to $S$ for processing, reducing the costs.
In the proposed offloading scheme, users can simultaneously transmit data on the same frequency, leading to higher computational efficiency and lower average cost than the benchmark.
As $w_1$ increases, the proportion of energy cost in the total cost increases and the proportion of delay cost in the total cost decreases.
Moreover, the average cost decreases slightly as $\omega _1$ increases because the delay cost is higher than the energy cost when ${\omega _1} = 0.5$, resulting in the total cost being more sensitive to the delay cost.
Thus, by adjusting each user's weighted factor ${\omega _1}$, a trade-off can be made between power consumption and delay.

\begin{figure}[t]
	\centering
	\subfigure[{3D trajectory}]{
		\label{3D}
		\includegraphics[width = 0.35 \textwidth]{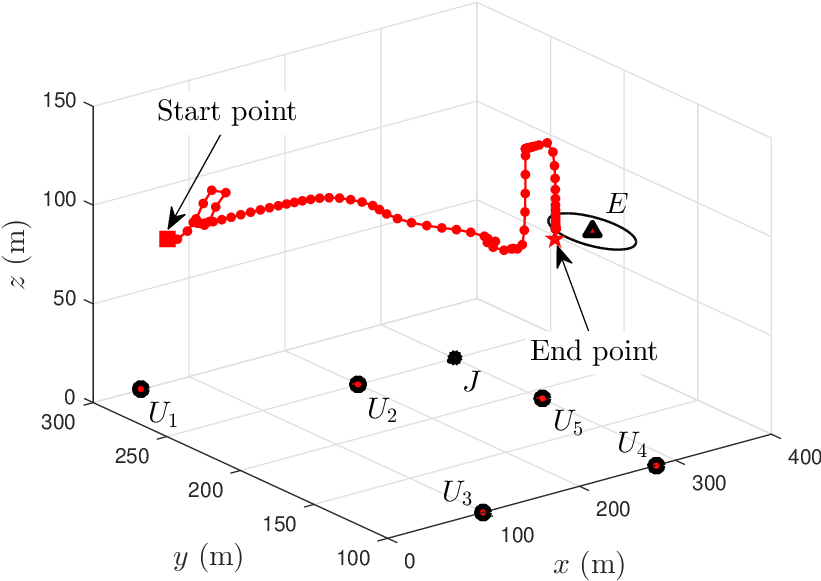}}
	\subfigure[{2D trajectory}]{
		\label{2D}
		\includegraphics[width = 0.35 \textwidth]{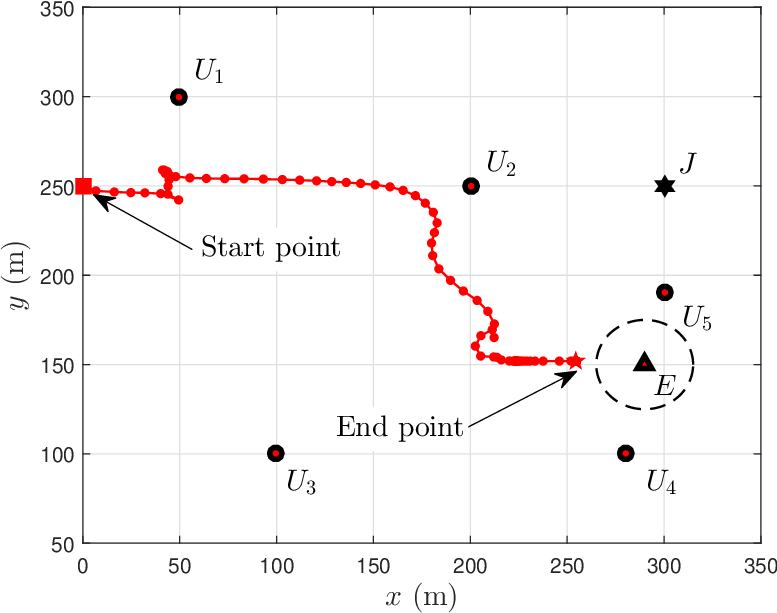}}
	\caption{{The optimized trajectory.}}
	\label{5trajectory}
\end{figure}

{
Fig. \ref{5trajectory} presents the optimized trajectory of $S$.
From the overall movement trend, it can be seen that $S$ first flies in the horizontal plane and approaches the users to improve the overall offloading rate.
To obtain a lower computation cost, $S$ flies to $U_2$, who is closer to $J$.
Due to the NOMA transmission mode, there is no need to worry about $U_3$ being left out.
To process the data of $U_4$ and $U_5$ who are close to $E$, $S$ chooses a compromise position and increases its altitude to enhance the elevation angle, thereby improving the channel gain of the $S$-$U_4$ and $S$-$U_5$ link.
Additionally, different from predetermined trajectories or goal-oriented trajectories in Ref. \cite{LuW2022TCOM}, the trajectory design in this work considers the average system cost, resulting in a more convoluted path.
}

\begin{figure}[t]
	\centering
	\includegraphics[width = 0.35 \textwidth]{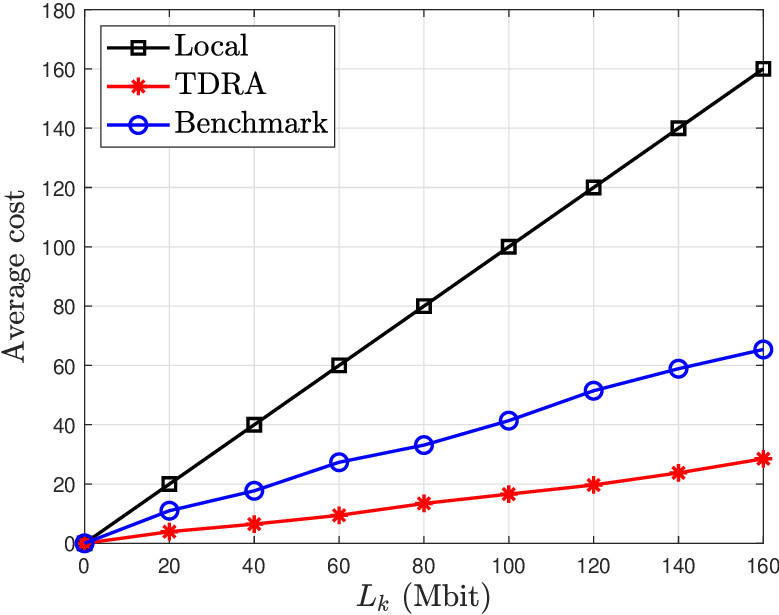}
	\caption{The average cost with varying $L_k$.}
	\label{6Lkcost}
\end{figure}

Fig. \ref{6Lkcost} illustrates user cost versus varying $L_k$.
Since the data that each user can process in each slot is constant, the energy consumption and latency incurred in each slot are also stable. As user data increases, power consumption and delay increase linearly, resulting in a linear rise in user cost.
As can be seen from the figure, the costs of all three schemes increase to different degrees as $L_k$ increases, and the larger the $L_k$ is, the larger gap between the cost of fully local computing and the cost of offloading computing.
In the lower-$L_k$ region,  the local device has the ability to process most of the data by itself, and only a small portion of the data will be offloaded to $S$ for processing, in this case, the difference between the proposed scheme and benchmark is not obvious, and the user cost is basically equal.
As increasing $L_k$, more data should be offloaded, the cost of the proposed scheme was reduced, which proves the efficiency of the proposed offloading scheme.

\begin{figure}[t]
	\centering
	\includegraphics[width = 0.35 \textwidth]{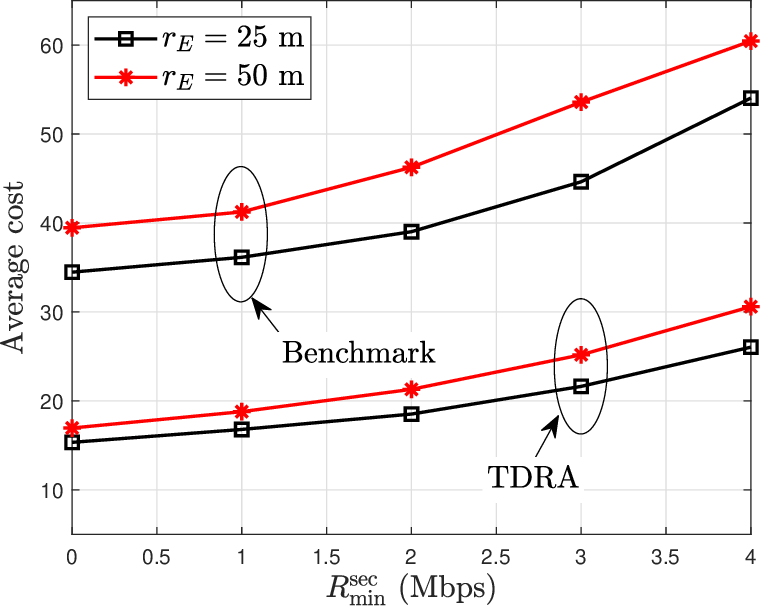}
	\caption{The average cost with varying $R_{\min }^{\sec}$ and $r_E$.}
	\label{7rEcost}
\end{figure}

Fig. \ref{7rEcost} describes the average cost with varying $R_{\min }^{\sec}$ and $r_E$.
One can observe that as $R_{\min }^{\sec}$ increases, the average cost also increases. This is attributed to the high $R_{\min }^{\sec}$, which denotes more stringent condition for user offloading, resulting in a decrease of the data offloaded by users, with more data being processed locally.
Moreover, larger $r_E$ results in large cost.
This is because larger $r_E$ signifies the higher uncertainty of the eavesdropper information in the considered system.

\section{Conclusion}
\label{sec:Conclusions}

In this work, security offloading in a NOMA-based multi-user UAV-MEC system has been investigated.
A novel optimization problem was formulated involving the concurrent design of the 3D trajectory, user transmission power, and CPU frequency to minimize the average computational cost.
A DDPG-based TDRA scheme was proposed to address the optimization challenge posed by high-dimensional continuous action space control in this scenario, aiming to achieve efficient user resource allocation and flight trajectory design of the UAV.
The simulation results confirmed that the proposed DDPG-based TDRA algorithm can effectively solve the secure offloading challenges posed by uncertain eavesdropping locations.
{Considering the scenarios with multiple UAVs and multiple terrestrial random-roaming or directional-walking users will be part of future work.}

\end{document}